\documentclass[11pt]{article}
\usepackage{amssymb, amsmath, amsthm, bm, float, graphicx,  tabularx, geometry, mathrsfs, setspace, enumerate, appendix, csquotes, url, yfonts, sectsty, eulervm, xcolor, hyperref, pdfpages}
\usepackage[T1]{fontenc}
\usepackage{xcolor}
\usepackage[round]{natbib}
\usepackage{pgf}
\usepackage{tikz}
\usepackage[round]{natbib}
\usetikzlibrary{arrows, automata, positioning}

\begin{document}

\title{Monetary policy and the racial wage gap}
\author{Edmond Berisha\thanks{Feliciano School of Business, Montclair State University, Montclair, NJ 07043; E-mail: berishae@montclair.edu}\and Ram Sewak Dubey\thanks{Feliciano School of Business, Montclair State University, Montclair, NJ 07043; E-mail: dubeyr@montclair.edu} \and Eric Olson\thanks{Collins College of Business, The University of Tulsa, Tulsa, OK 74104; Email: edo4695@utulsa.edu}
}
\date{\today}
\maketitle

\begin{abstract}
\noindent This paper aims to clarify the relationship between monetary policy shocks and  wage inequality.
We emphasize the relevance of \emph{within} and \emph{between} wage group inequalities in explaining total wage inequality in the United States. 
Relying on the quarterly data for the period 2000-2020, our analysis shows that racial disparities explain 12\% of observed total wage inequality. 
Subsequently, we examine the role of monetary policy in wage inequality. 
We do not find compelling evidence that shows that monetary policy plays a role in exacerbating the racial wage gap. 
However, there is evidence that accommodative monetary policy plays a role in magnifying between group wage inequalities but the impact occurs after 2008. 

\noindent \emph{Keywords:}\;\texttt{Decomposition,} \; \texttt{Monetary Policy,}\; \texttt{Inequality.}

\noindent \emph{JEL}: \texttt{D60,}\; \texttt{O40,}\; \texttt{O50.}
\end{abstract}

\setcounter{MaxMatrixCols}{10}

\newpage

\section{Introduction}\label{s1} 
\indent It is well-known that the level of economic disparity has considerably increased in the last three decades in most developed countries, particularly in the United States (US). One of the observed characteristics of the social and economic disparity has been the differences in wages between races. 
Figures 1, 2, and 3 show the wage gap across Asians, African Americans, and Whites within each wage group (first decile, third quartile, and ninth decile) for the 2000-2020 time period.  As can be seen, racial wage gaps within the first decile group remained relatively stable over the sample period. Specifically, Whites and Asians made 40\% and 60\% more than African Americans over the entire sample period. However, interesting trends appear as one examines the upper part of the wage distribution. 
Within the third quartile, the wage gap between Whites and African Americans was relatively steady with Whites making 30\% more than African Americans. Asians, on the other hand, experienced the largest wage growth over the last 20 years. Initially, Asians made 40\% more than African Americans but that gap increased to 70\% by the end of our sample. Similar trends are observed for the three racial groups in the ninth decile. Again, Asians saw the largest wage growth relative to the other two racial groups over the last two decades. 

The literature identifies several sources for the observed racial disparities in earnings. 
First, the departure of manufacturing plants and union jobs from urban areas in the Northeast and Midwest (\textcolor{blue}{\citet{mccall2001sources}}). 
Second, international trade and immigration may affect the wage gap due if they increase the supply of low-skilled workers and reduce wages and employment of low- skilled natives (see \textcolor{blue}{\citet{borjas1996searching}}, \textcolor{blue}{\citet{card2001immigrant}}, and \textcolor{blue}{\citet{berisha2018income}}). 
However, relatively little is known in the literature regarding the role monetary policy may play as a potential driver of the racial wage gap. 
While previous research has documented the role monetary policy may have in increasing non-labor income inequality (stocks, dividends etc.), scrutiny of the role that monetary policy plays in labor income is not present in the current literature. 
Several studies document that contractionary monetary policy shocks increase income inequality \textcolor{blue}{(\citet{areosa2016}},\textcolor{blue}{\citet{furceri2018}}, \textcolor{blue}{\citet{coibion2017})}. \textcolor{blue}{\citet{bartscher2021monetary}}, on the other hand, points to a lack of sufficient evidence to conclude that accommodative monetary policy plays a significant role in reducing racial inequities.

Several Federal Reserve Officials have made recent arguments suggesting that central banks need to set-up policies that promote a more inclusive economy that leads to lower racial inequalities.%
\footnote{Interested reader is referred to \href{https://www.atlantafed.org/about/atlantafed/officers/executive_office/bostic-raphael/message-from-the-president/2020/06/12/bostic-a-moral-and-economic-imperative-to-end-racism.aspx}{www.atlantafed.org}. for an exposition of current thought of the Fed on the subject.} As such, our aim in this study is to examine the impact that conventional and unconventional monetary policy actions of the Federal Reserve have had on wage inequality measures. 
Previous literature focuses on the dynamic responses of indices that summarizes inequality into a single numerical index.%
\footnote{Note, in calculating inequality measure these studies use income, which is the overall money one gets. 
Our study relied on the wage income for calculating the inequality index.} 
We argue that analyzing distributional consequences of monetary policy using a single numerical index fails to capture responses of inequality across different races or other subsets within the wage distribution. 
For example, if one examines the bottom quantile of the distribution, dynamics of a single numerical index will not clarify the distributional consequences of monetary policy within that specific quantile. 
We posit that a wage inequality measure that allows us to examine inequality within various subsets is required to provide a clear lens through which we could analyze distributional effects of monetary policy decisions. 
We construct an additively decomposable inequality measure (Theil Index) that incorporates wages of Asian, African Americans and White Americans who work full-time from three points in the wage distribution.%
\footnote{As suggested by the referee, we have checked data on other points in the distribution and are able to confirm that our analysis and conclusions remain unaltered.}
We subsequently use exogenous US monetary policy shocks (in conventional and unconventional time periods) to examine monetary policy affects. 

Consistent with previous literature, we find a persistent increase in overall wage inequality in the US over our sample period (Q1:2000 to Q1:2020). Interestingly, we find that most of the observed wage inequality in the US is driven by increases in wage inequality within each race, not increases in wage inequality between races. Our measures suggest that at most, 12\% of overall wage inequality is attributed to the racial wage gap. We also contribute to the relatively new literature by documenting that monetary policy shocks is a statistically significant driver of wage inequality. We find evidence that expansionary monetary policy shocks contributed to higher wage inequality in the post-2008 time-period. However, our findings highlight that the response of \textit{between} group inequality within each race is the main driver of the relationship between monetary policy shocks and total wage inequality. We find little statistical evidence to show that monetary policy shocks matter for wage differences between races. Our results suggest that fiscal policy tools would be better suited in tackling racial wage gap. The remainder of the paper is organized as follows. In section \ref{s2}, we outline the construction and decomposition of the Theil Index. Section \ref{s3} describes the data and the methodology used to analyze the impact of monetary policy shocks on the evolution of  wage inequality measure that incorporates racial wage gap. Section \ref{s4} presents our results.
We conclude in section \ref{s5}.
 
\section{Wage Inequality Decomposition}\label{s2}
\subsection{Theil Measures and its decomposability}\label{s24}

Let $\mathbb{N}$, and $\mathbb{R}$  denote the set of natural numbers and real numbers, respectively.
For $n\in \mathbb{N}$, $\mathbb{R}^n$ denote the set of all vectors having $n$ real components.
We denote the null vector with $n$ entries,  $(0, \cdots, 0)$, by $0^n$ and the vector with all entries equaling one $(1, \cdots, 1)$ by $u^n$.
For $x$, $y\in \mathbb{R}^n$, we say $x\geq y$ if $x_i\geq y_i$ for $i=1, \cdots, n$; $x>y$ if $x\geq y$ and $x\neq y$; $x\gg y$ if $x_i>y_i$ for $i=1, \cdots, n$.
For $x\geq 0^n$, we denote $\sum_{i=1}^n x_i$, the sum of all coordinates, by $|x|$ and $\frac{|x|}{n}$, the mean, by $\overline{x}$.
We use notation $(x; y)$ to denote the combination vector $\left(x_1, \cdots, x_n, y_1, \cdots, y_m\right)$ where $x\in \mathbb{R}^n$ and $y\in \mathbb{R}^m$.
A partition of vector $y\in \mathbb{R}^n$ into $l\geq 2$ smaller vectors is denoted by $\left(y^{(1)}, \cdots, y^{(l)}\right)$ where $y^{(1)}$, $y^{(l)}$ are the sub-vectors. 

Following \textcolor{blue}{\citet{foster1983}}, we introduce the notion of inequality measure which permits both the population size and total wage to be variable.
For a given population size $n\geq 1$, consider the set $D^n := \left\{ x\in \mathbb{R}^n: x>0^n\right\}$ of $n$-wage distributions.
Each $n$-distribution $y$ specifies a scheme of allocating total wage of $|y|$ among a population consisting of $n$ persons.
The term inequality index is used to denote a function $I^n:  D^n \rightarrow \mathbb{R}$ which describes comparisons of inequality for any given level of population $n$.
The inequality measure provides comparison of inequality across different population sizes.
It is defined on $\mathscr{D} := \overset{\infty}{\underset{n=1}{\bigcup}} D^n$, the set of all distributions.%
\footnote{Put differently, while inequality index takes the population size as given and fixed, the inequality measure is defined over all possible population sizes and wage distributions.} The inequality measure of our interest is the Theil measure introduced in \textcolor{blue}{\citet{theil1967}}.
The Theil measure is  defined as $I: \mathscr{D}\rightarrow \mathbb{R}$, whose indices are defined as follows:
\begin{equation}\label{e1}
I^n(y) =\sum_{i=1}^{n} \frac{y_i}{|y|} \ln \left(n \frac{y_i}{|y|}\right),  y\in D^n.
\end{equation}
It is easy to infer that the least value for the Theil index is zero, indicating perfect equality. 
Further, the upper bound of the index is the natural logarithm of the sample size and is therefore the index  $I^n$ could increase unboundedly with the sample size $n$. \textcolor{blue}{\citet{theil1967}} demonstrated that in a sample comprising of multiple groups of wage generators, the inequality measure $I^n (y)$, as defined in (\ref{e1}), can be expressed by the sum of two separate terms. 
The first term, \enquote{within group} inequality, is defined as the sum of inequality levels of each group weighted by the share of wage it generates as a proportion of the total wage of the sample. 
The second term, \enquote{between-group} inequality, captures the inequality in a \enquote{smoothed} wage distribution where the mean wage of the group is treated as the wage of the group.%
\footnote{For example, if $y=(1,3,5,7,3)$, and $y$ is partitioned into two sets $y^{(1)}= (1, 3)$  and $y^{(2)} = (5, 7, 3)$. 
The decomposability property implies that $I(y) = \left(\frac{4}{19}\right) I(y^{(1)}) + \left(\frac{15}{19}\right)I(y^{(2)})+ I(2, 2, 5, 5, 5)$.}
Following \citet{foster1983}, we express the decomposition  in formal terms as follows.
Let the aggregate population of size $n$ be partitioned in $l\geq 2$ groups having $n_1$, $\cdots$, $n_l$ (with $n_1+\cdots+n_l=n$) members respectively. 
Let the wage distributions of the smaller groups be $y^{(i)} \in D_{n_{i}}$ ($i= 1, \cdots, l$).
Then the Theil inequality index $I^n(y)$ can be decomposed as 
\begin{equation}\label{e2}
I(y)= \sum_{i=1}^l \left(\frac{|y^{(i)}|}{|y|} I(y^{(i)})\right) + I(\overline{y}^{(1)} u^{n_1}; \overline{y}^{(2)} u^{n_2};\cdots; \overline{y}^{(l)}u^{n_l}),
\end{equation}							  
where $y = \left(y^{(1)}; \cdots; y^{(l)}\right)$.

We use the above method to calculate the Theil Index that presents overall wage inequality in the US that incorporates wages of three racial groups (Asians, African Americans, and Whites) across the three subsets (first decile, third quartile, and ninth decile). Then, we decompose it into within-and-between wage groups. Note, the \textit{within} component captures the racial gap amongst different races within each wage group, whereas, the \textit{between} component captures the gap between wage groups in the wage distribution. The analysis allow us to identify how racial disparities based on wage hold at the bottom, middle, and top of wage distribution.

\section{Data and methodology}\label{s3}

We use data from the Bureau of Labor Statistics' (BLS) \footnote{https://www.bls.gov/news.release/pdf/wkyeng.pdf} Current Population Survey (CPS).
CPS provides basic information on the labor force, employment, and unemployment.%
\footnote{The survey is conducted monthly for the BLS by the U.S. Census Bureau using a scientifically selected national sample of about 60,000 eligible households, with coverage in all 50 states and the District of Columbia.} The earnings data are collected from one-fourth of the CPS monthly sample and are limited to wage and salary workers. All self-employed workers, both incorporated and unincorporated, are \emph{excluded} from the CPS earnings estimates. To construct the wage disparity measure and subsequently decompose it into within and between groups disparities, we use the wages at the first decile, the third quartile, and the ninth decile of wage distribution for the three racial groups recognized by U.S. Census (Whites, Black or African American, and Asians) for the period 2020-Q1 to 2021-Q1.%
\footnote{Sample selection is driven by the availability of the  data at quarterly frequency.}
Weekly wages obtained from the CPS are in current dollars for workers employed full time and older than 16 years. 
The wage measure represents earnings before taxes and other deductions and include any overtime pay, commissions, or tips usually received (at the main job in the case of multiple jobholders). 

\subsection{Racial wage inequality: Within and between groups}\label{s31}

Figures 4, 5, and 6 report the wage inequality for the US calculated from (\ref{e1}) and its Theil decomposition. 
Figure 4 displays the overall wage inequality over the sample period.
It shows that wage inequality in the U.S. increased persistently from the early 2000s to 2020Q1. 
Figure 5 displays the contributions of inequality within the first decile, third quartile, and ninth decile to the overall wage inequality reported in Figure 4. 
Put another way, the time series in Figure 5 show how inequality amongst Whites, Asians, and African Americans have evolved within each point in the wage distribution. 
As can be seen in Figure 4, there has been an increase in the racial wage gaps for those in the ninth decile (red line) as well as the third quartile (blue line). 
However, note that the racial wage gap has not materially changed for individuals in the first decile of the wage distribution. 
While there has been an uptick in the racial wage gaps in the ninth decile and third quartile, it should be noted that the racial wage gap accounts for approximately 12\% (8\%+4\%) of total wage inequality observed in Figure 4. 

Figure 6 displays the contribution of inequality \textit{between} the  three subsets to overall wage inequality seen in Figure 4. Interestingly, note that while it does trend downwards, in Figure 6, 88\% to 94\% of overall wage inequality in US is attributed to wage inequality between subsets. Similar evidences are observed in Figures 13, 14, and 15, where we calculate the wage growth rates for the three wage groups among three races (Asians, African Americans, and Whites). The speed of growth in wages increases the further up you go in wage distribution. Thus, our decomposition suggests that much of the changes in wage inequality over the last 20 years has been a result of differences between subsets of the wage distribution rather than between racial groups. 
Our results are consistent with \citet{goldin2007}, and \citet{acemoglu2011} that education likely plays a key role in the growth in earnings inequality (assuming that those in the ninth decile have more education than those in the first).

\subsection{Methodology}
To capture the causal impact of monetary policy changes on wage inequality measures and its within-and-between components requires using \enquote{exogenous} monetary policy changes \textcolor{blue}{(\citet{furceri2018})}. As such, we use the \textcolor{blue}{\citet{bu2021unified}} quarterly monetary policy shock series, which are the sum of monthly orthogonalized movements of zero-coupon yields with maturities of 1 year to 30 years. \textcolor{blue}{\citet{bu2021unified}} develop a heteroscedasticity-based, partial least squares (PLS) approach to identify US monetary policy shocks. 
The series exhibit relatively large movements, which makes it interesting to explore the response of wage inequality measure to fluctuations in these monetary policy shocks. 
Figure 7 displays the \textcolor{blue}{\citet{bu2021unified}} monetary shock time series. 

We use a VAR to quantify the overall effects of monetary policy shocks on total wage inequality and its within-and-between components. That is, 
\begin{equation}\label{e3}
\textbf{X}_t = \textbf{A}_0 + \textbf{B}(\textbf{L}) \textbf{X}_{t-l} + \textbf{C}(\textbf{L}) d_{t-l} + \textbf{e}_{t}
\end{equation}
where $\textbf{X}_t$ is a vector that contains the inequality measures, $d_t$ is the \textcolor{blue}{\cite{bu2021unified}} monetary policy shock series. 
Note, in model (\ref{e3}), we are imposing the restriction that exogenous monetary policy shocks do not respond to changes in our wage inequality measures within the quarter, which seems reasonable. 
We estimate the impulse responses of wage inequality and its components over a 10-quarter period to a unit shock in monetary policy series. 
We include one period lagged of inequality measures to control for persistence that exists in inequality series and 4 lags of monetary policy shocks $d_t$. 
For robustness, we also include a measure of output, \emph{industrial production}, in $\textbf{X}_t$ to control for dynamics in economic conditions.   

\section{Results}\label{s4}
\subsection{Impulse Responses} \label{s44} 

We first regressed our monetary shock series on the overall inequality measure reported in Figure 4. 
Figure 8 displays the impulse responses from the VAR as well as one-standard deviation confidence bands. Note that an expansionary monetary policy shock initially decreases wage inequality but after 3 quarters increases wage inequality by 0.5 standard deviations. Figures 9 and 10 display the impulse responses of monetary policy shocks on wage inequality before and after the 2008 financial crisis. Note, that the effect remains statistically significant only for the period after the Great Recession (Q1:2009 to Q1:2020) with point estimates which are roughly twice as large as those pre-2008. As such, our results suggest that monetary policy actions of the Federal Reserve to support the recovery after the Great Recession likely exacerbated wage inequality in the US rather than mitigating it. For robustness, we also included industrial production in model (3) to control for overall business conditions in US. As expected (see Figure 11), expansionary monetary shocks lead to higher wage inequality as well as higher industrial production. Specifically, a 25 basis point cut in monetary policy leads to 0.6 standard deviation increase in industrial production and 0.6 standard deviation increase in the Theil index.  

Figure 12 shows impulse responses of within-and-between components of Theil Index calculated from our VAR in (3). Again, the within components correspond to the racial wage gaps whereas the between components correspond to the inequality between wages at different deciles of the wage distribution. A couple of points are worth highlighting. First, we can see that expansionary monetary policy shocks decrease the racial wage inequality in the first decile and the third quartile after 11 quarters; however, the effect is close to zero and not statistically significant at conventional levels. Expansionary monetary policy shocks increase wage inequality in the ninth decile (0.24 standard deviation) and the between group component (0.4 standard deviation). However, again, it should be noted that the results are only statistically significant for the between group panel. That is, monetary policy shocks only effect the inequality between groups within each race. 

In summary, our results suggest that expansionary monetary policy shocks slightly decrease the racial wage gap in the first decile and third quartile, albeit the results are not statistically different from zero. Expansionary monetary policy shocks increase racial the wage gap in the ninth decile (but not statistically different from zero) and increase the wage gap between the different wage distribution points.

\subsection{Discussion} \label{s54} 

Endogenous growth theory may provide insight into our results. Consider the canonical model with two sectors that use different types of capital goods as inputs (physical and human capital), and a Cobb-Douglas ($Y=(K^{\alpha}, E\cdot L^{\alpha})$ production function that exhibits constant returns to scale in each sector. If the sector that is more physically capital intensive (and assuming that the ninth decile group derives most of their wage from the human capital sector relative to the first decile and third quartile) is also less interest-rate elastic, expansionary monetary shocks would increase wage inequality between the two sectors if the distribution of workers is different across industries. That is, investment would increase more in the sector that uses more human capital relative to physical capital thereby increasing the relative wage gaps between the two sectors.  
For example, note in the Figure below the two investment demand curves ($\text{Inv}_{1,d}$ and $\text{Inv}_{2,d}$) where investment is simply the change in the capital stock for each respective industry.
\includegraphics{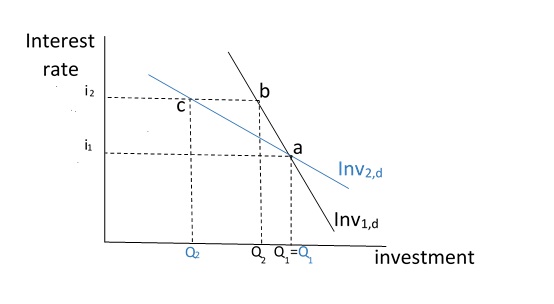}

Let $\text{Inv}_{2,d}$ denote the investment demand curve for the industry that uses relatively more human capital and $\text{Inv}_{1,d}$ denote the investment demand curve for the industry that use relatively more physical capital. 
An expansionary monetary shock that moves interest rates from $i_2$ to $i_1$ increases investment in the relatively human capital intensive sector from point $c$ to $a$ but only increases investment in the relatively physical capital intensive sector from $b$ to $a$. 
Thus, given no change in the labor supply in each industry, capital and wage per effective worker will increase more in the industry that is more interest-rate elastic thereby increasing wage disparities. This type of explanation is consistent with the observed. Figures 13, 14, and 15 show the wage growth rates for the three wage groups among three races (Asians, African Americans, and Whites). As you go up in wage distribution, the speed of growth in wages increases within the three racial groups. It should be noted that this mainly happens during the period of post Great Recession until 2016, while the Federal Reserve kept the interest rate close to zero. Thus, wages of workers at upper end of wage distribution among the three races are more sensitive to monetary policy changes relative to wages of workers at lower end of wage distribution.

\section{Conclusions}\label{s5}

Our results are notable in two ways. First, we empirically document the relevance of racial-and between wage groups' inequalities in explaining total wage inequality using the Theil index. We show that the importance of racial wage gap in the total wage inequality increases towards the end of the sample period (2015 to 2019). The racial wage gap explains at most 12\% of total wage inequality. Second, we contribute to the relatively new literature by documenting that expansionary monetary policy shocks, post 2008, exacerbated wage inequality in the US. Results reveal that the response of between groups inequality within each race is the main driver of the relationship between monetary policy shocks and total wage inequality. 


\newpage

\bibliographystyle{plainnat}
\bibliography{APAnonymity}
\newpage
\begin{figure}[htp]
    \centering
    \includegraphics[width=0.85\textwidth]{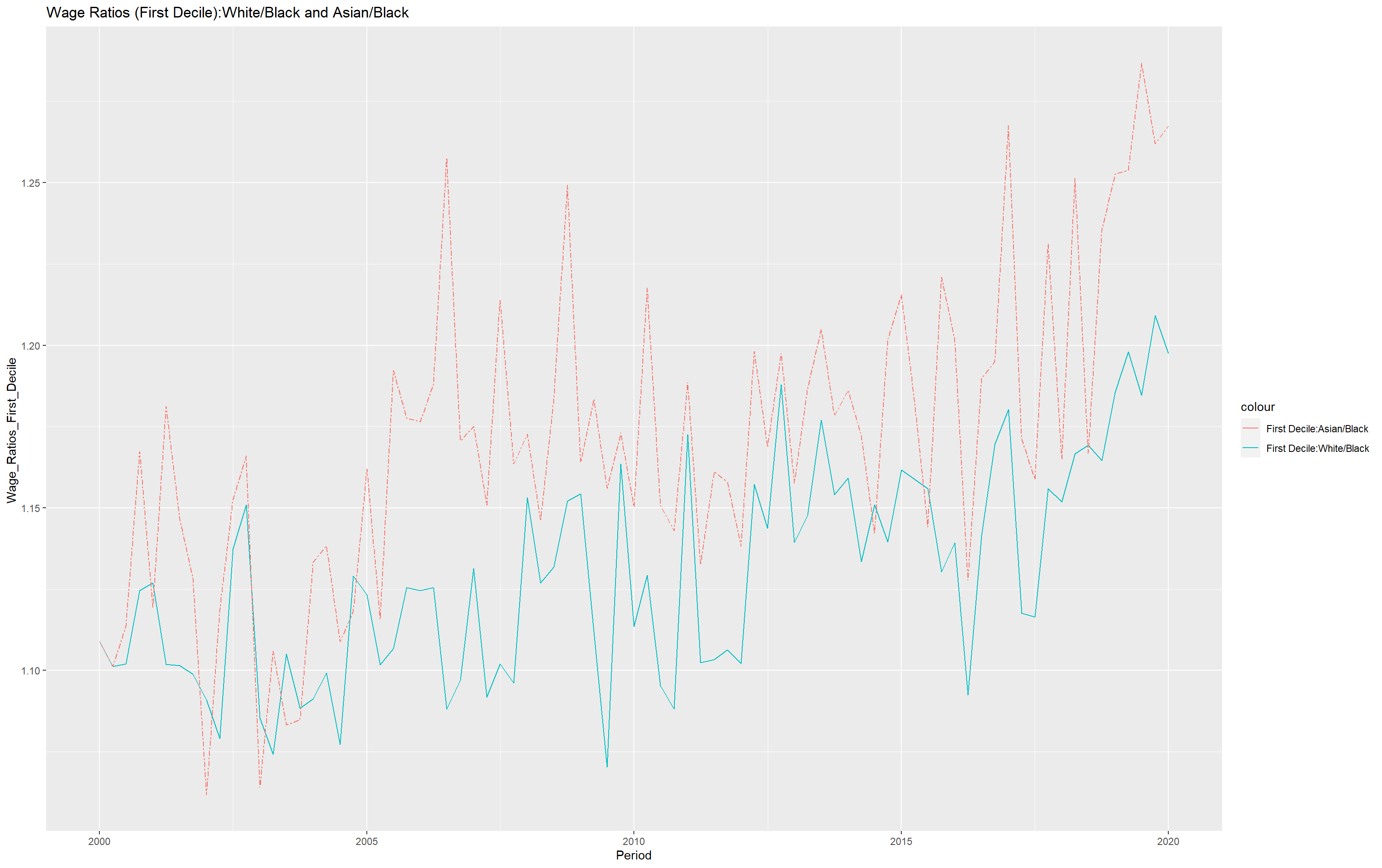}
    \caption{Racial Wage Gap within First Decile}
    \label{top}
\end{figure}

\begin{figure}[htp]
    \centering
    \includegraphics[width=0.85\textwidth]{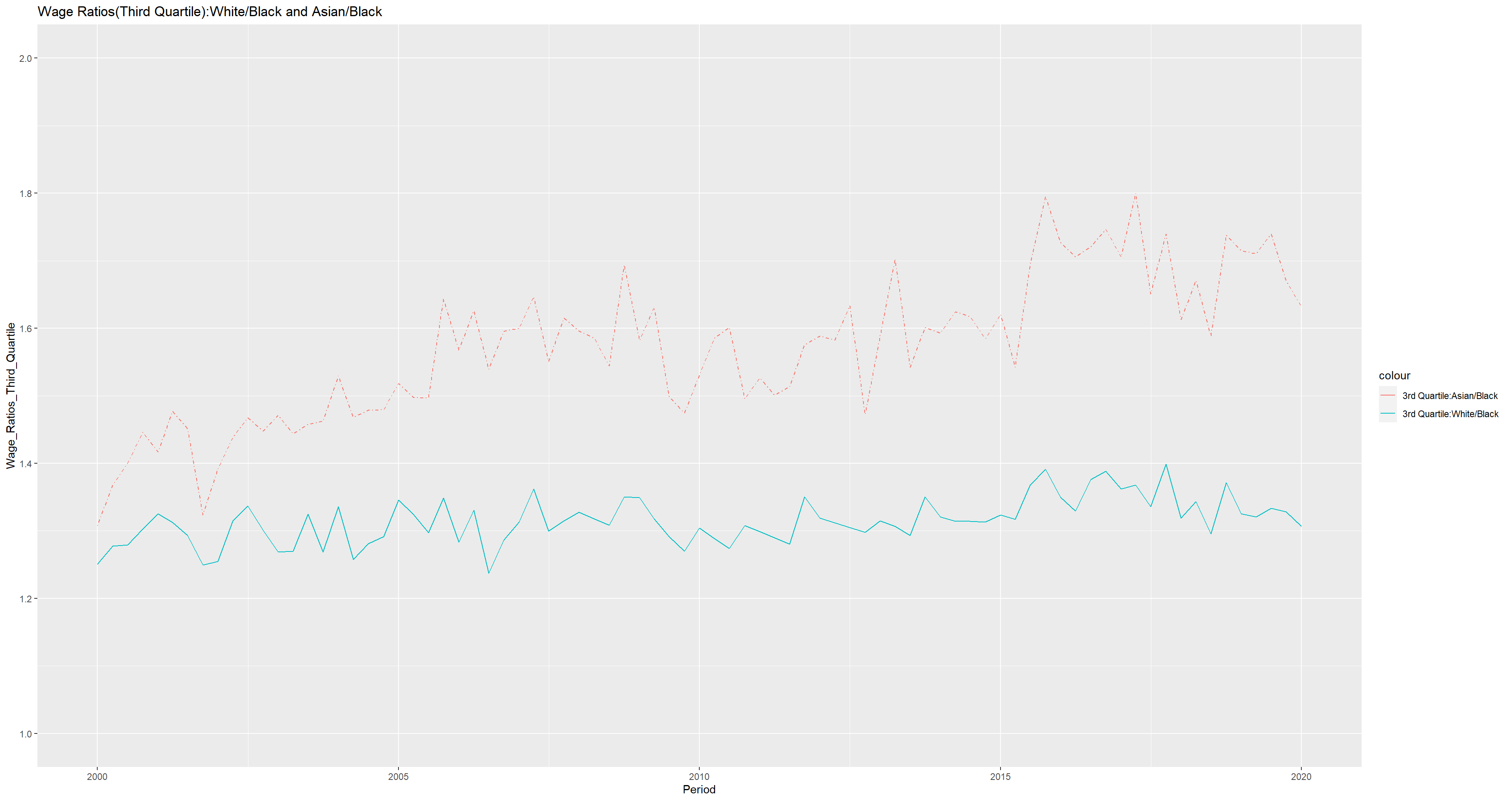}
    \caption{Racial Wage Gap within Third Quartile}
    \label{fig:Pre-Tax National Income Shares}
\end{figure}

\begin{figure}[htp]
    \centering
    \includegraphics[width=0.85\textwidth]{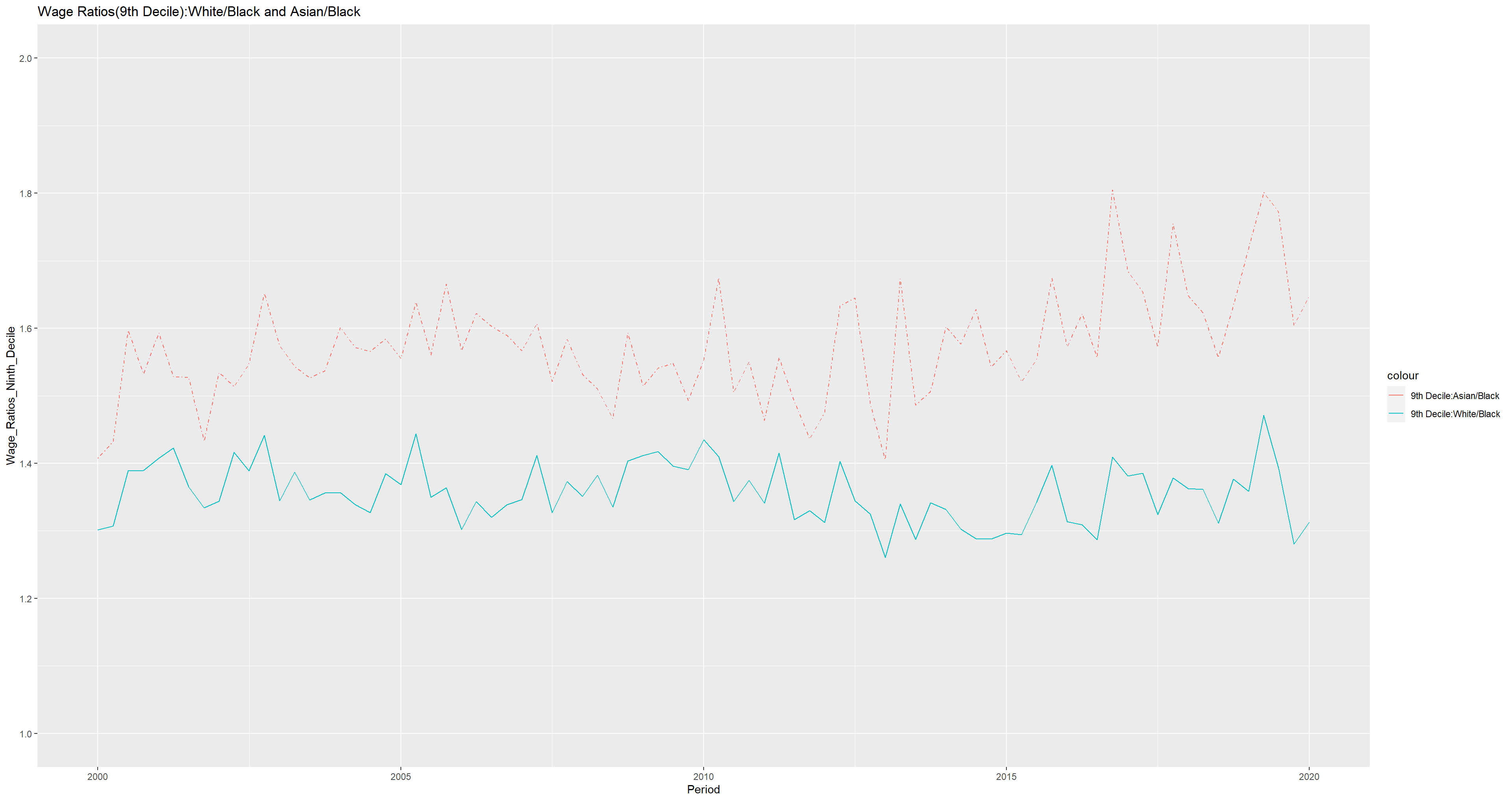}
    \caption{Racial Wage Gap within Ninth Decile}
    \label{fig:Pre-Tax National Income Shares}
\end{figure}

\newpage
\begin{figure}[htp]
    \centering
    \includegraphics[width=0.85\textwidth]{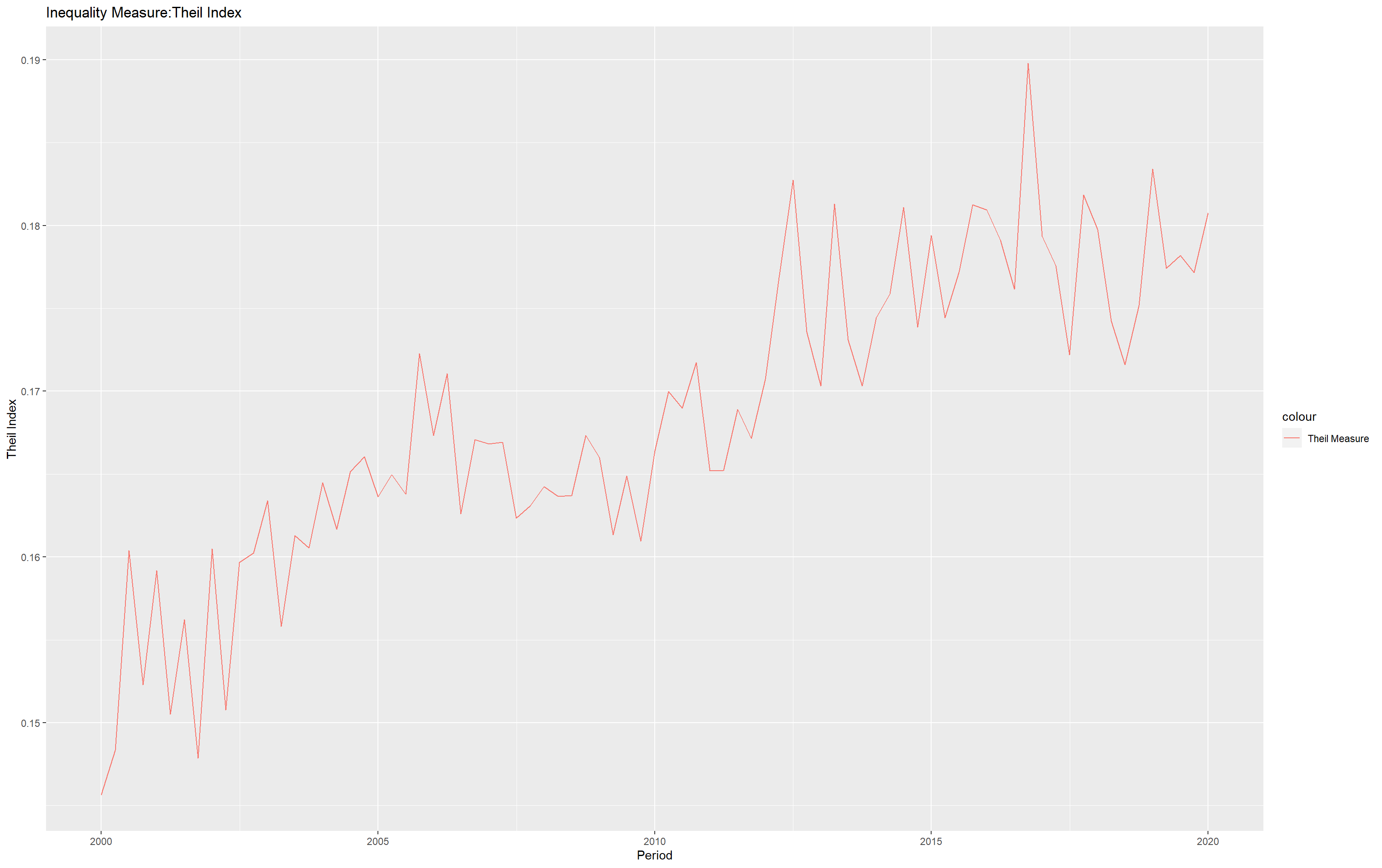}
    \caption{Wage Inequality Measure: Theil Index}
    \label{top}
\end{figure}

\begin{figure}[htp]
    \centering
    \includegraphics[width=0.85\textwidth]{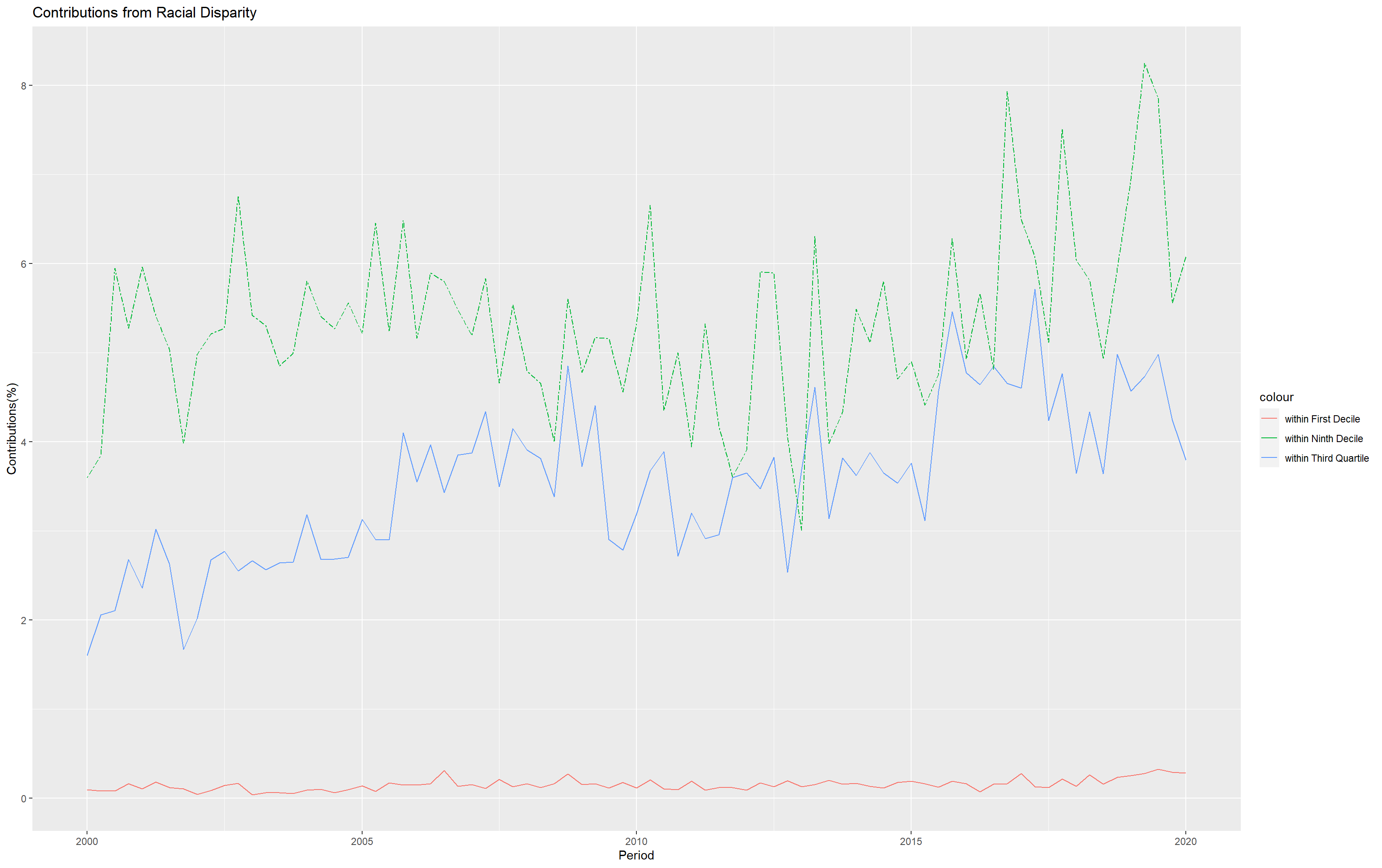}
    \caption{Contribution from Racial Disparity}
    \label{fig:Pre-Tax National Income Shares}
\end{figure}

\newpage
\begin{figure}[htp]
    \centering
    \includegraphics[width=0.85\textwidth]{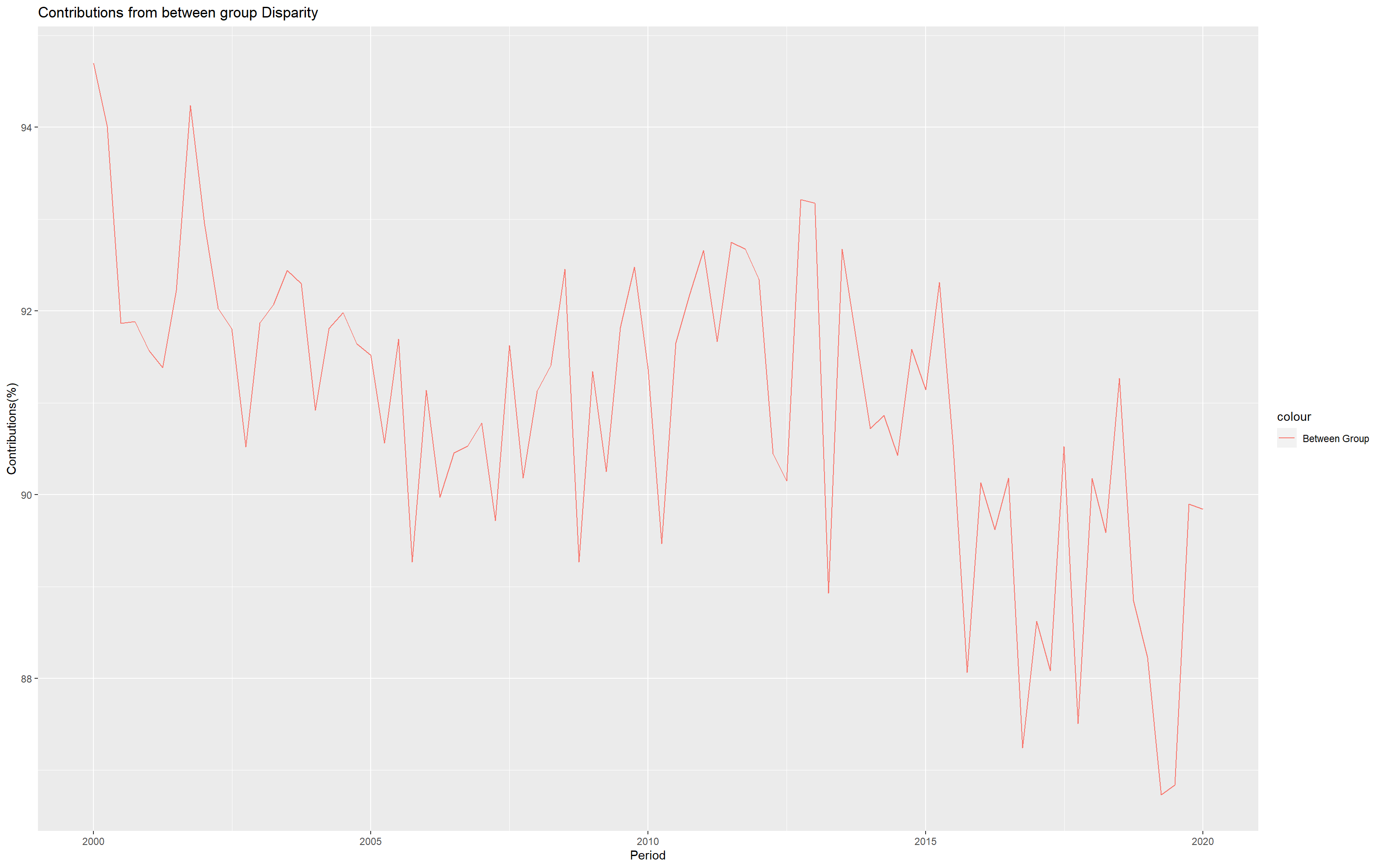}
    \caption{Contribution from Between Group Disparity}
    \label{top1}
\end{figure}

\begin{figure}[htp]
    \centering
    \includegraphics[width=0.85\textwidth]{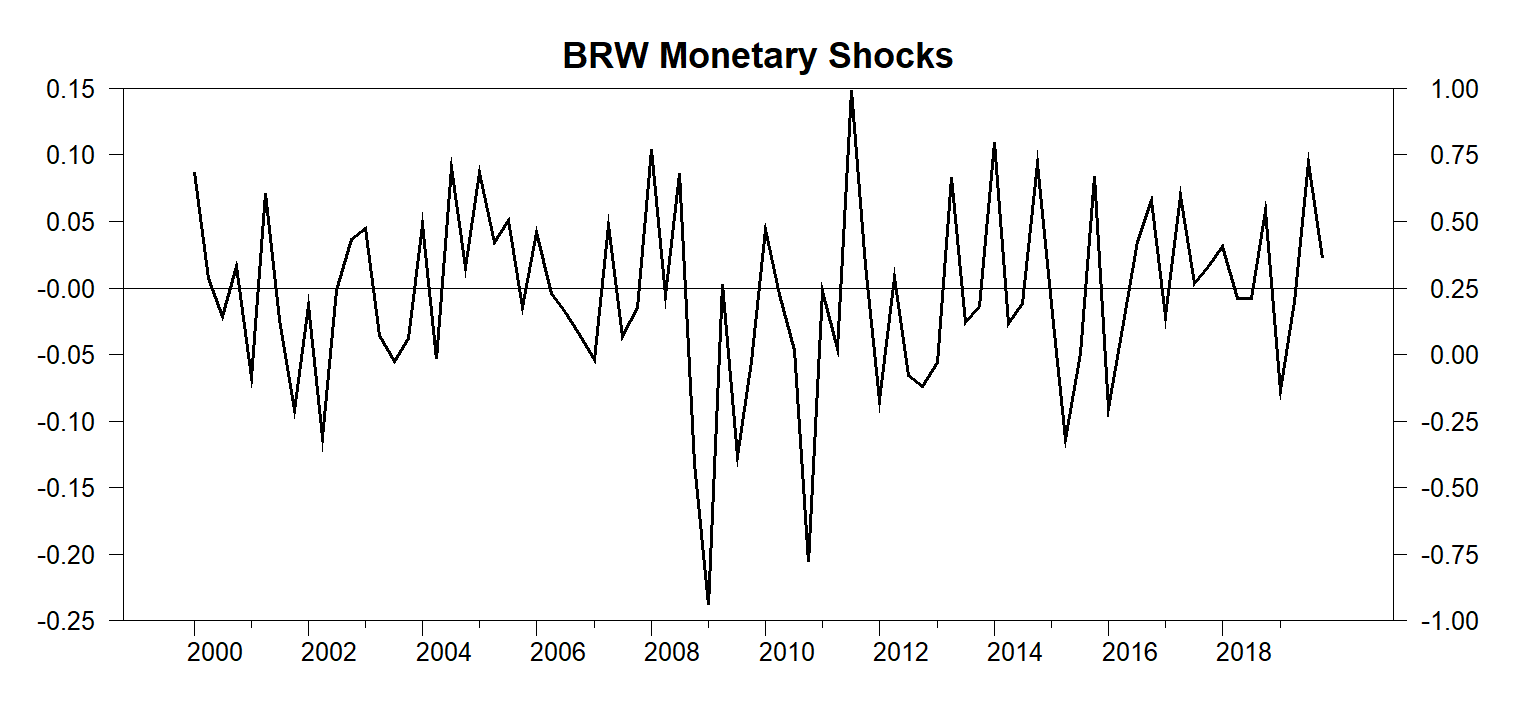}
    \caption{BRW Monetary Shocks}
    \label{top2}
\end{figure}

\newpage

\begin{figure}[htp]
    \centering
    \includegraphics[width=0.85\textwidth]{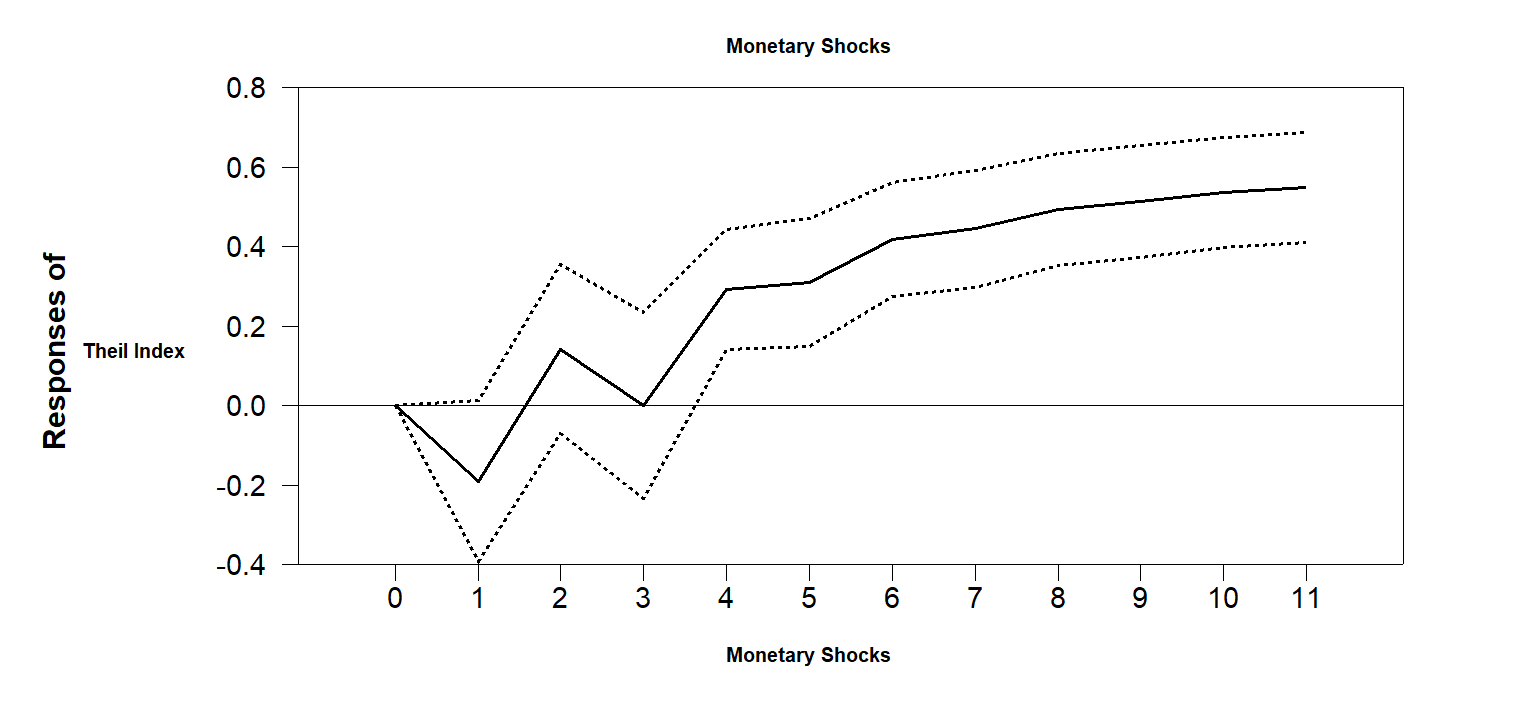}
    \caption{Response of Theil Index to a 25 basis points drop in Monetary Policy variable}
    \label{top3}
\end{figure}

\begin{figure}[htp]
    \centering
    \includegraphics[width=0.85\textwidth]{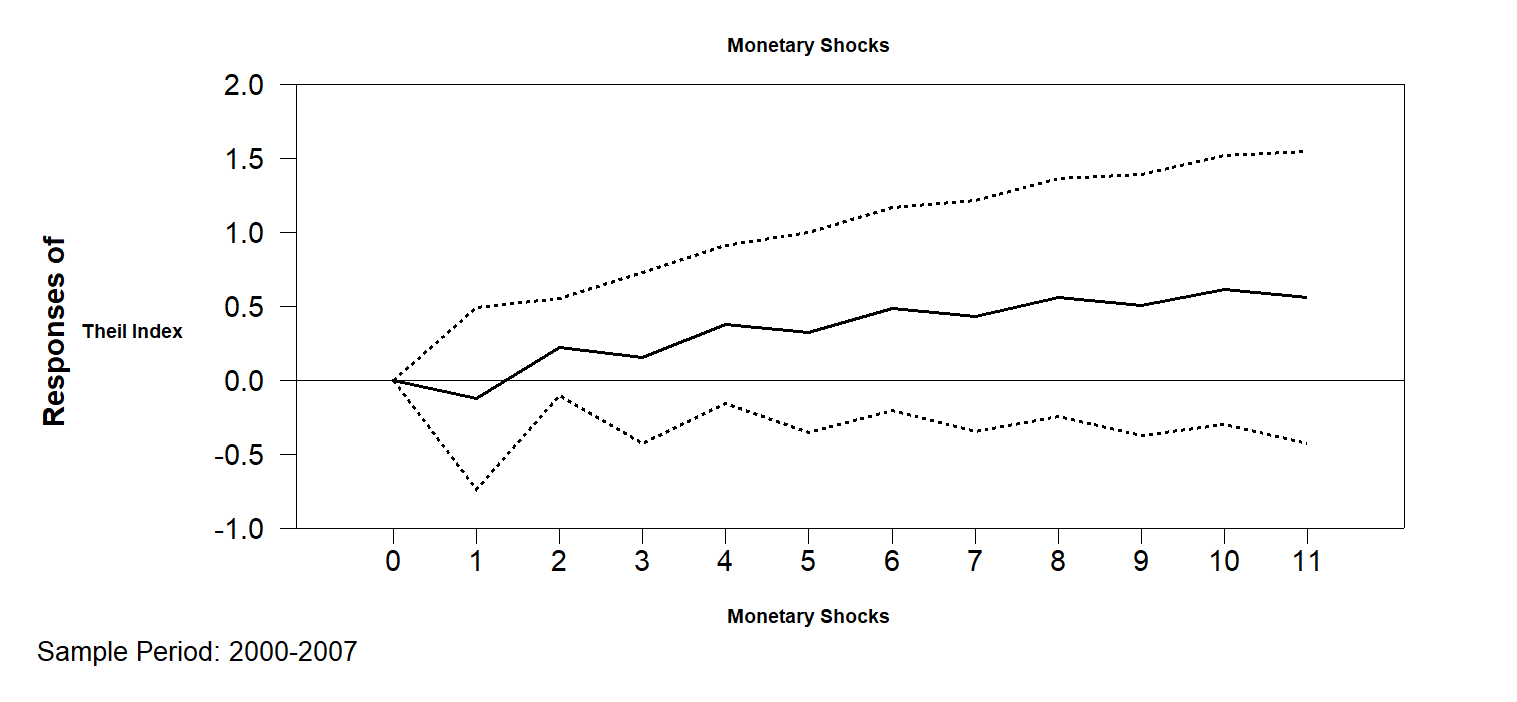}
    \caption{Response of Theil Index to a 25 basis points drop in Monetary Policy variable: 2000 to 2007}
    \label{fig:Pre-Tax National Income Shares1}
\end{figure}

\begin{figure}[htp]
    \centering
    \includegraphics[width=0.85\textwidth]{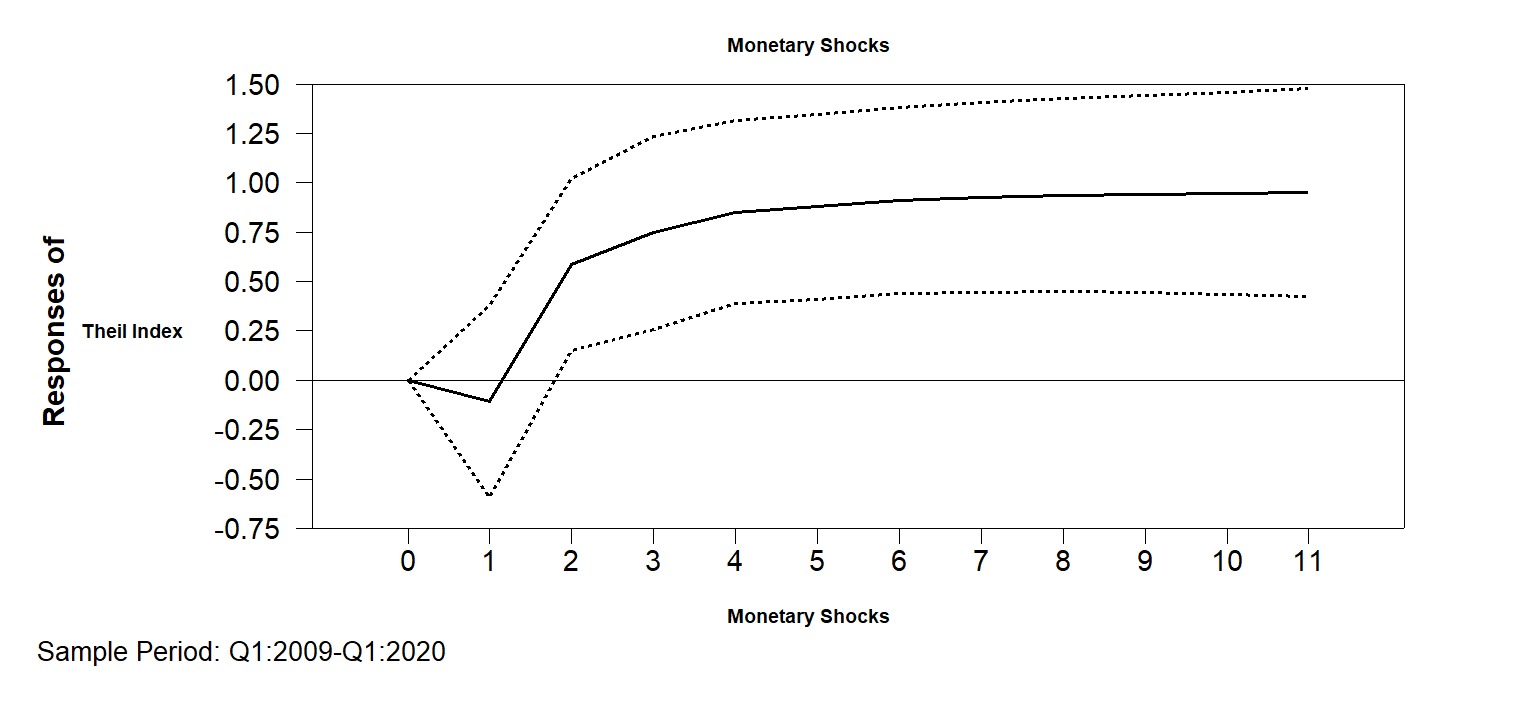}
    \caption{Response of Theil Index to a 25 basis points drop in Monetary Policy variable: 2009 to 2020}
    \label{top4}
\end{figure}

\begin{figure}[htp]
    \centering
    \includegraphics[width=0.85\textwidth]{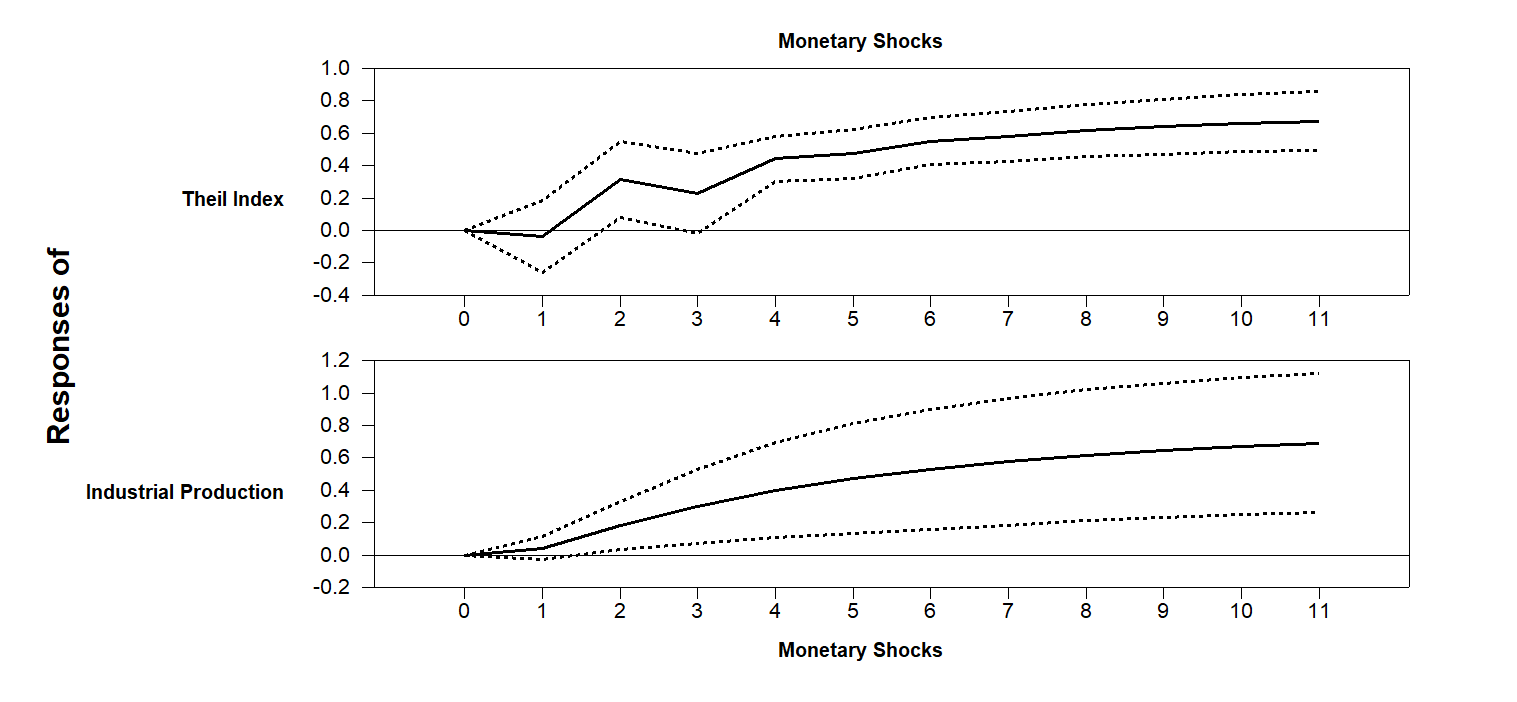}
    \caption{Response of Theil Index to a 25 basis points drop in Monetary Policy variable: after Controlling for Industrial Production}
    \label{fig:Pre-Tax National Income Shares2}
\end{figure}

\begin{figure}[htp]
    \centering
    \includegraphics[width=0.85\textwidth]{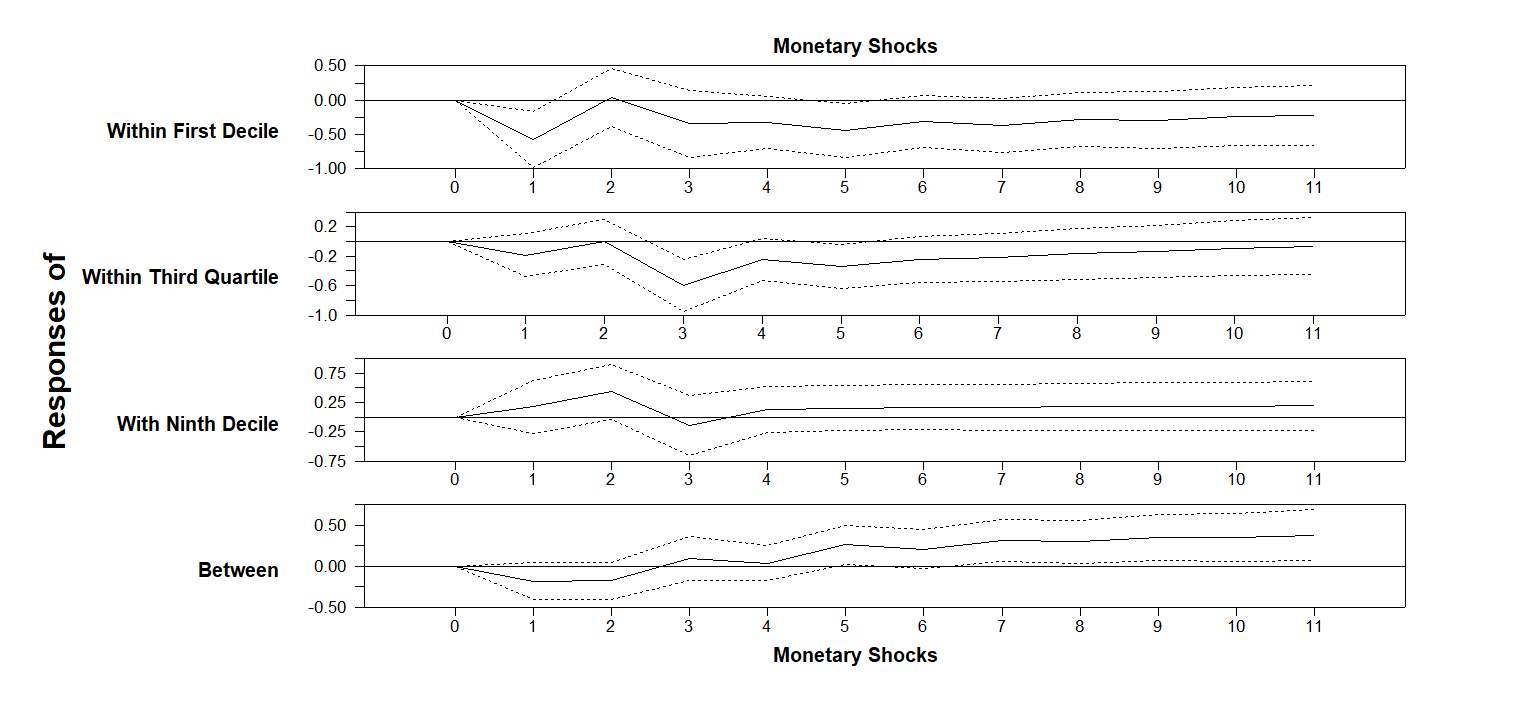}
    \caption{Response of Within and Between components to a 25 basis points drop in Monetary Policy variable}
    \label{fig:Pre-Tax National Income Shares3}
\end{figure}

\begin{figure}[htp]
    \centering
    \includegraphics[width=0.85\textwidth]{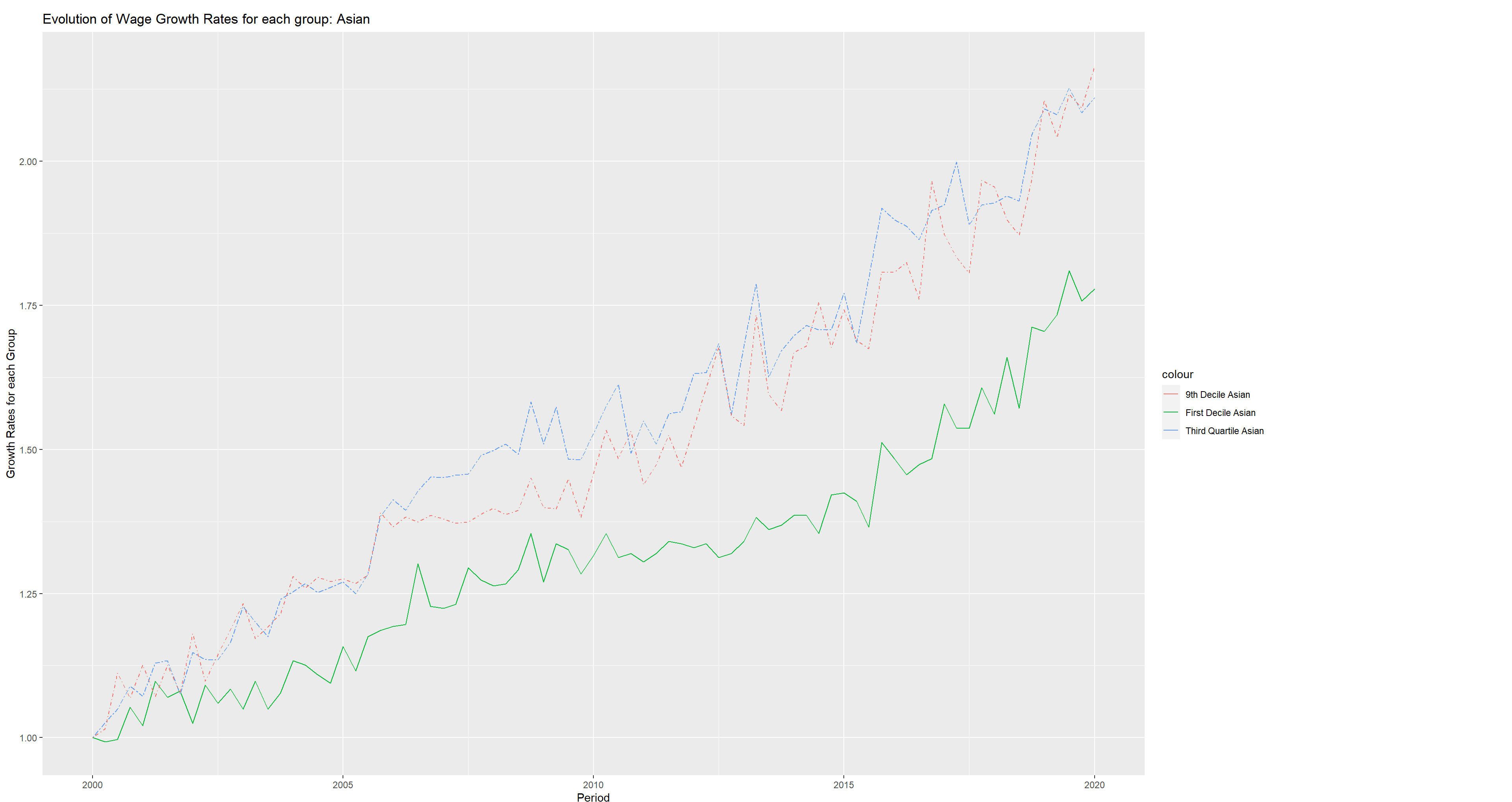}
    \caption{Wage Growth rate across three wage groups: Asian}
    \label{fig:Pre-Tax National Income Shares4}
\end{figure}

\begin{figure}[htp]
    \centering
    \includegraphics[width=0.85\textwidth]{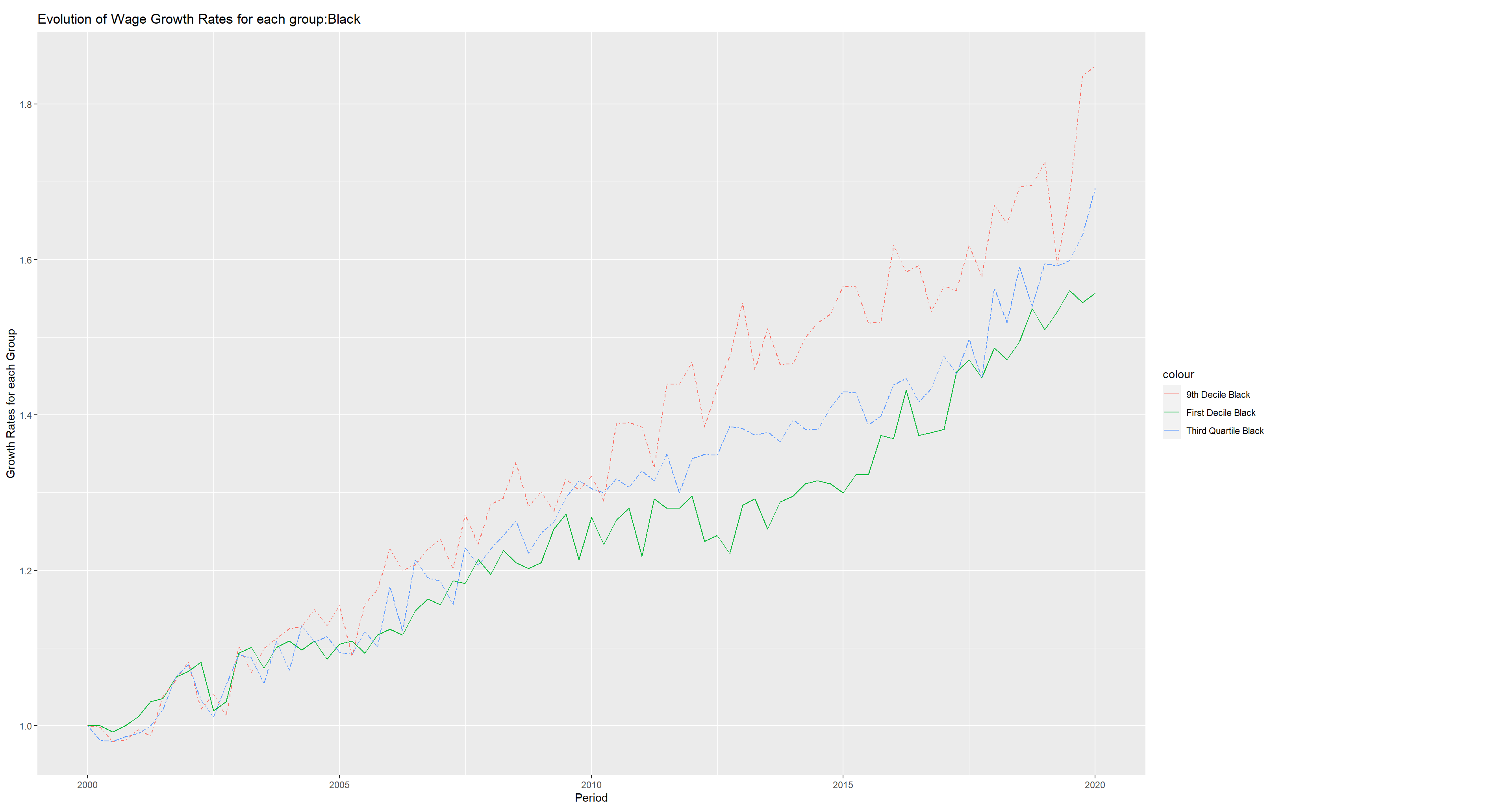}
    \caption{Wage Growth rate across three wage groups: Black}
    \label{fig:Pre-Tax National Income Shares5}
\end{figure}
    
\begin{figure}[htp]
    \centering
    \includegraphics[width=0.85\textwidth]{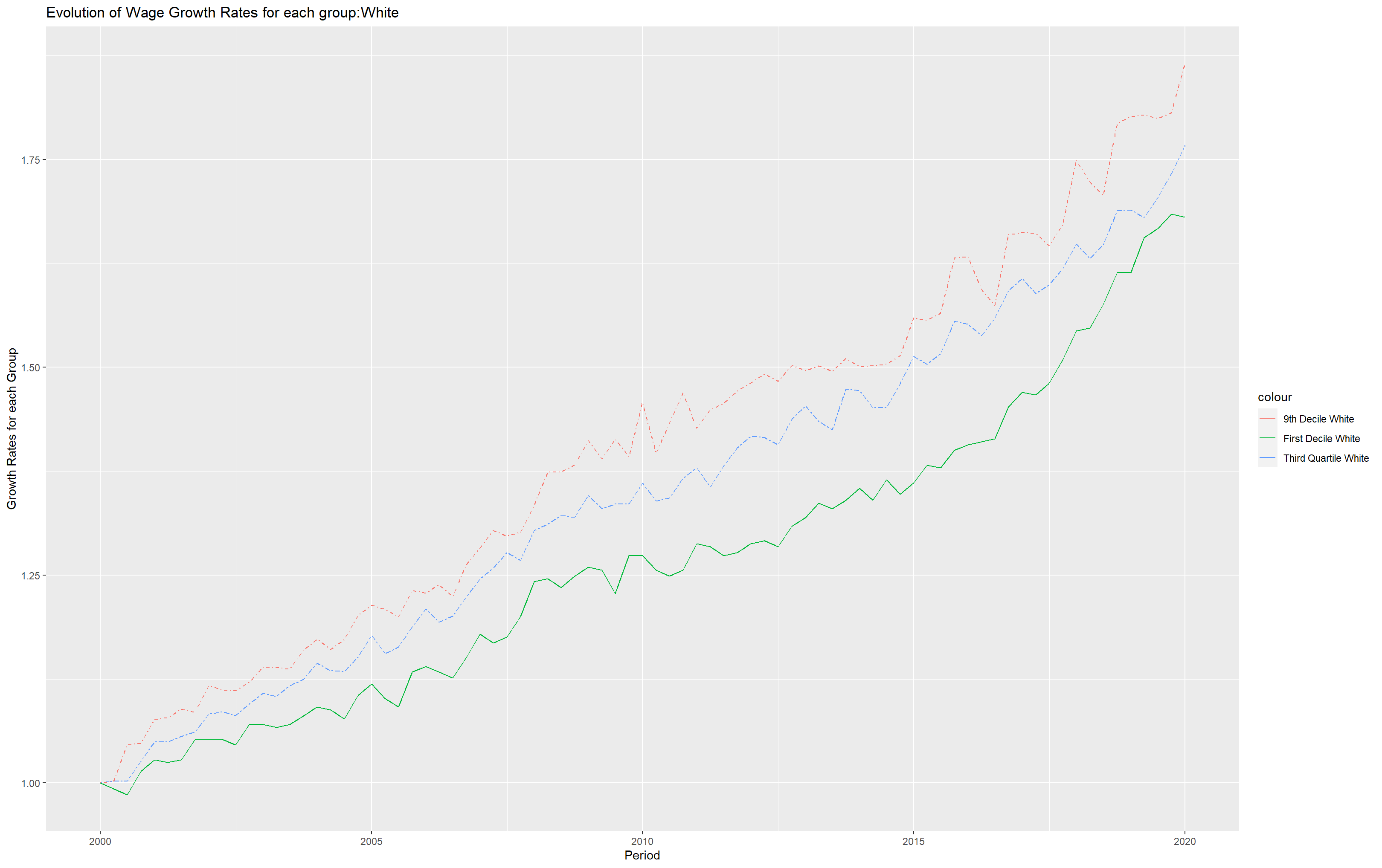}
    \caption{Wage Growth rate across three wage groups: White}
    \label{fig:Pre-Tax National Income Shares6}
\end{figure}

\end{document}